\documentclass[useAMS,usecolumn,usenatbib,usegraphicx]{mn2e}

\title[Giant Ly$\alpha$ blobs at $z=3$]{The Subaru Ly$\alpha$ blob survey: A sample of 100 kpc Ly$\alpha$ blobs at $z=3$\thanks{Based on data collected at Subaru Telescope, which is operated by the National Astronomical Observatory of Japan.}}
\author[Y. Matsuda et al.]{
\parbox[t]{\textwidth}{\vspace{-1cm}
Y. Matsuda,$^{\! 1}$\thanks{E-mail: yuichi.matsuda@durham.ac.uk} T. Yamada,$^{\! 2}$ T. Hayashino,$^{\! 3}$ R. Yamauchi,$^{\! 3}$ Y. Nakamura,$^{\! 2, 3}$ N. Morimoto,$^{\! 2}$ M. Ouchi,$^{\! 4,5}$\thanks{Carnegie fellow.} Y. Ono,$^{\! 6}$ K. Kousai,$^{\! 3}$ E. Nakamura,$^{\! 3}$ M. Horie,$^{\! 3}$ T. Fujii,$^{\! 3}$ M. Umemura,$^{\! 7}$ and M. Mori\,$^{\! 7}$}\\\\
$^{1}$Department of Physics, Science Site, Durham University, South Road, Durham, DH1 3LE\\
$^{2}$Astronomical Institute, Graduate School of Science, Tohoku University, Aramaki, Aoba-ku, Sendai 980-8578, Japan\\
$^{3}$Research Center for Neutrino Science, Graduate School of Science, Tohoku University, Sendai 980-8578, Japan\\
$^{4}$Observatories of the Carnegie Institution of Washington, 813 Santa Barbara Street, Pasadena, CA 91101, USA\\
$^{5}$Institute for Cosmic Ray Research, University of Tokyo, Kashiwa 277-8582, Japan\\
$^{6}$Department of Astronomy, Graduate School of Science, The University of Tokyo, Tokyo 113-0033, Japan\\
$^{7}$Center for Computational Sciences, University of Tsukuba, Tsukuba 305-8577, Japan}
\begin{document}

\date{Accepted ... ; Received ... ; in original form ...}

\pagerange{\pageref{firstpage}--\pageref{lastpage}} \pubyear{2010}

\maketitle

\label{firstpage}

\begin{abstract}
We present results of a survey for giant Ly$\alpha$ nebulae (LABs) at $z=3$ with Subaru/Suprime-Cam. We obtained Ly$\alpha$ imaging at $z=3.09 \pm 0.03$ around the SSA22 protocluster and in several blank fields. The total survey area is $2.1$ square degrees, corresponding to a comoving volume of $1.6 \times 10^6$ Mpc$^3$. Using a uniform detection threshold of $1.4 \times 10^{-18}$ erg s$^{-1}$ cm$^{-2}$ arcsec$^{-2}$ for the Ly$\alpha$ images, we construct a sample of 14 LAB candidates with major-axis diameters larger than $100$ kpc, including five previously known blobs and two known quasars. This survey triples the number of known LABs over 100 kpc. The giant LAB sample shows a possible "morphology-density relation": filamentary LABs reside in average density environments as derived from compact Ly$\alpha$ emitters, while circular LABs reside in both average density and overdense environments. Although it is hard to examine the formation mechanisms of LABs only from the Ly$\alpha$ morphologies, more filamentary LABs may relate to cold gas accretion from the surrounding inter-galactic medium (IGM) and more circular LABs may relate to large-scale gas outflows, which are driven by intense starbursts and/or by AGN activities. Our survey highlights the potential usefulness of giant LABs to investigate the interactions between galaxies and the surrounding IGM from the field to overdense environments at high-redshift.
\end{abstract}


\begin{table*}
 \centering
 \begin{minipage}{153mm}
  \caption{Summary of narrow-band observations}
  \begin{tabular}{@{}ccccccccc@{}}
  \hline
   Field     &   RA         &    Dec      &  Date &  Exposure & Area   & FWHM &  \multicolumn{2}{c}{Depth} \\
             &  (J2000)     &  (J2000)    &  (mm/yyyy) & (hours) & (arcmin$^2$)  & (arcsec) & (cgs)$^a$ & (ABmag)$^b$\\
 \hline
 SXDS-C & 02:18:00.0  &  $-$05:00:00  & 08, 09, 10/2005 & 5.2 & 682  & 1.0  & 0.81 & 26.3 \\
 SXDS-N & 02:18:00.0  &  $-$04:35:00  & 10/2005 & 4.8 & 740  & 1.0 & 0.94 & 26.2 \\
 SXDS-S & 02:18:00.0  &  $-$05:25:00  & 08, 10/2005 & 4.8 & 737  & 1.0 & 0.82 & 26.3 \\
 GOODS-N & 12:37:23.6  &  +62:11:31  & 04/2005 & 10.0 &  869 & 1.1 & 0.69 & 26.6 \\
 SDF & 13:24:39.0  &  +27:29:26  & 04/2004, 04/2005 & 7.2 &  805 & 1.0 & 0.67 & 26.5 \\
 SSA22-Sb1 & 22:17:34.0  &  +00:17:01  & 09/2002 & 7.2 & 633 & 1.0 & 0.92 & 26.3 \\
 SSA22-Sb2 & 22:16:36.7  &  +00:36:52  & 08/2004 & 5.5 &  487 & 1.0 & 0.96 & 26.3 \\
 SSA22-Sb3 & 22:18:36.3  &  +00:36:52  & 08, 09/2005 & 5.5 &  537 & 1.0 & 0.89  & 26.3\\
 SSA22-Sb4 & 22:19:40.0  &  +00:17:00  & 08, 09, 10/2005 & 5.5  &  529 & 1.1 & 1.15 & 25.9 \\
 SSA22-Sb5 & 22:15:28.0  &  +00:17:00  & 09/2005 & 5.5 &  565 & 1.0 & 1.06 & 26.1 \\
 SSA22-Sb6 & 22:14:30.7  &  +00:33:52  & 10/2005 & 5.5 &  572 & 1.0 & 0.92 & 26.3 \\
 SSA22-Sb7 & 22:17:42.7  &  +00:56:52  & 09, 10/2005 & 5.5 &  480 & 1.0 & 1.02 & 26.2 \\ 
\hline
\end{tabular}
$^a$The 1-$\sigma$ surface brightness limit ($10^{-18}$ erg s$^{-1}$ cm$^{-2}$ arcsec$^{-2}$).\\
$^b$The 5-$\sigma$ limiting magnitude calculated with 2 arcsec diameter aperture photometry.\\
\end{minipage}
\end{table*}

\section{Introduction}

Ly$\alpha$ blobs (LABs) are spatially extended Ly$\alpha$ nebulae seen in the high-redshift Universe \citep[e.g.,][]{1996ApJ...457..490F, 1999AJ....118.2547K, 2000ApJ...532..170S, 2004ApJ...602..545P, 2004AJ....128..569M, 2005ApJ...629..654D, 2006A&A...452L..23N, 2007MNRAS.382...48G, 2007MNRAS.378L..49S, 2009ApJ...702..554P, 2009ApJ...693.1579Y}. LABs are thought to relate to the formation of massive galaxies \citep{2005ApJ...629..654D, 2006ApJ...640L.123M} and to be indicative of strong interactions between the inter-galactic medium (IGM) and galaxies with intense star-formation activities and/or AGNs \citep{2005ApJ...622....7F}. To explain the formation mechanisms of LABs, at least three possible ideas have been proposed: cold gas accretion, galactic winds, and photoionization by central galaxies or by AGNs \citep{2000ApJ...537L...5H, 2000ApJ...532L..13T, 2001ApJ...548L..17C}. In spite of extensive observational and theoretical efforts in the decade after the first discovery of LABs, the formation mechanisms of LABs are still controversial \citep{2006Natur.440..644M, 2009ApJ...700....1G, 2009MNRAS.400.1109D, 2010MNRAS.tmp..933G, 2010arXiv1005.3041F, 2010MNRAS.406..913S}. 

Among the LABs, special attention have been given to the largest examples with the spatial extents of $\sim 100-200$ kpc (hereafter giant LABs) because of their spectacular morphologies and possible association with protoclusters \citep{2000ApJ...532..170S, 2004ApJ...602..545P, 2008ApJ...678L..77P, 2009MNRAS.400L..66M}. At present, there are only six known giant LABs over 100 kpc, and they have been selected by using different quality data sets and different methods \citep{1996ApJ...457..490F, 2000ApJ...532..170S, 2005ApJ...629..654D, 2007MNRAS.382...48G,  2009MNRAS.400L..66M}. It is therefore difficult to examine their statistical properties. In order to construct a statistically reliable sample of giant LABs and so test their possible association with overdense environments, we undertook a deep, wide-field Ly$\alpha$ imaging at $z=3.1$. 

In this letter, we use AB magnitudes and adopt cosmological parameters, $\Omega_{\rm M} = 0.3$, $\Omega_{\Lambda} = 0.7$ and $H_0 = 70$ km s$^{-1}$ Mpc$^{-1}$. In this cosmology, the Universe at $z=3.1$ is 2.0 Gyr old and $1.0$ arcsec corresponds to a physical length of 7.6 kpc at $z=3.1$.

\section[]{Observations and Data Reduction}

We briefly describe our observations and data reduction, although the details will be reported in a separate paper (Yamada et al. in prep). The summary of the observations and data is listed in Table~1. The imaging observations were carried out between September 2002 and October 2005 using Suprime-Cam \citep{2002PASJ...54..833M} on the 8.2-m Subaru Telescope \citep{2004PASJ...56..381I}. Suprime-Cam has a pixel scale of $0.2$ arcsec and a field of view of $34 \times 27$ arcmin$^2$. We obtained narrow-band ($NB497$) images for 12 pointings: Great Observatories Origins Deep Survey-North (GOODS-N), Subaru Deep Field (SDF), three fields in Subaru-XMM Deep Survey (SXDS-C, N, and S) and seven fields around SSA22a (SSA22-Sb1-7). The SSA22-Sb1 field was the first field of our survey and centred at SSA22a, which contains the protocluster region at $z=3.09$ discovered by \citet{2000ApJ...532..170S}. Initial results of the observations in the SSA22-Sb1 have been already published \citep{2004AJ....128.2073H, 2004AJ....128..569M, 2005ApJ...634L.125M, 2006ApJ...640L.123M}. The $NB497$ filter has a central wavelength of 4977 \AA\ and FWHM of 77 \AA\, which corresponds to the redshift range for Ly$\alpha$ at $z=3.062-3.126$  \citep{2004AJ....128.2073H}. The width of the redshift slice is 59 comoving Mpc. For the SSA22 fields, we obtained broad-band ($B$ and $V$) images in our observing runs. For GOODS-N, we used archival raw $B$ and $V$-band images \citep{2004AJ....127..180C}. For the SDF and SXDS fields, we used public, reduced $B$ and $V$-band images \citep{2004PASJ...56.1011K, 2008ApJS..176....1F}.

\begin{figure}
\centering
\includegraphics[scale=0.65]{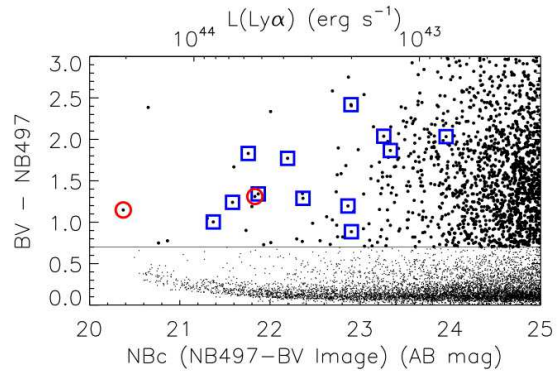}
  \caption{Colour magnitude plot of $BV - NB497$ vs. $NB_{\rm c}$ for $NB_{\rm c}$ detected sources (black dots). The solid line represents colour criterion ($BV-NB497=0.7$) used for narrow-band excess objects (larger dots). The blue squares and red circles indicate giant (major-axis diameters$\ge 100$ kpc) LAB candidates without QSO and with QSO, respectively. All magnitudes and colours are measured with isophotal apertures.}
\end{figure}

\begin{table*}
 \centering
 \begin{minipage}{180mm}
  \caption{Properties of the 14 giant LAB candidates}
  \begin{tabular}{@{}cccccccccc@{}}
  \hline
 ID     &   RA         &    Dec      & a$^a$ &  Area  &  $L_{\rm Ly\alpha}$  & $F^b$ & $\delta_{\rm LAE}$ & $z_{\rm spec}$ & Note \\
  &  (J2000) & (J2000) & (kpc) & (arcsec$^2$) & ($10^{43}$ erg s$^{-1}$) & & &  & \\
 \hline 
SSA22-Sb1-LAB1 & 22:17:25.95 & +00:12:37.7 & $175$ & $181\pm14$ & $ 8.1\pm 0.6$ & 0.56 & 2.7 & 3.099$^c$ & 8$\mu$m$^i$/submm$^j$ \\
SSA22-Sb6-LAB1 & 22:13:48.30 & +00:31:32.8 & $166$ & $116\pm9$ & $ 5.8\pm 0.4$ & 0.69 & 0.6 & 3.094$^d$ &  --- \\
SSA22-Sb1-LAB2 & 22:17:38.99 & +00:13:27.8 & $157$ & $137\pm8$ & $ 6.8\pm 0.3$ & 0.59 & 3.7 & 3.091$^c$  & X-ray$^k$/8$\mu$m$^i$ \\
SSA22-Sb5-LAB1 & 22:15:33.56 & +00:25:16.9 & $147$ & $59\pm7$ & $ 3.8\pm 0.4$ & 0.80 & -0.5 & --- &  --- \\
SSA22-Sb3-LAB1 & 22:17:59.45 & +00:30:55.7 & $126$ & $102\pm8$ & $20.4\pm 0.3$ & 0.52 & 1.2 & 3.099$^e$ & QSO$^e$/Radio$^l$ \\
GOODS-N-LAB1 & 12:35:57.54 & +62:10:24.9 & $124$ & $47\pm7$ & $ 5.4\pm 0.5$ & 0.77 & 0.9 & 3.075$^f$ & QSO$^f$/X-ray$^m$ \\
SSA22-Sb2-LAB1 & 22:16:58.37 & +00:34:32.0 & $121$ & $60\pm15$ & $ 2.0\pm 0.6$ & 0.70 & 1.2 & --- & --- \\
SSA22-Sb2-LAB2 & 22:16:56.40 & +00:27:53.3 & $115$ & $48\pm11$ & $ 1.4\pm 0.2$ & 0.73 & -0.1 & --- & --- \\
SSA22-Sb1-LAB5 & 22:17:11.66 & +00:16:44.4 & $110$ & $43\pm11$ & $ 1.3\pm 0.3$ & 0.74 & 1.0 & ---  & 8$\mu$m$^i$/submm$^n$ \\
SSA22-Sb5-LAB2 & 22:15:30.27 & +00:27:43.6 & $107$  & $53\pm7$ & $ 2.1\pm 0.3$ & 0.66 & -0.1 & ---  & --- \\
SSA22-Sb6-LAB4 & 22:14:09.58 & +00:40:54.6 & $107$ & $32\pm4$ &$ 2.0\pm 0.2$ & 0.79 & -0.1 & 3.116$^d$ & --- \\
SSA22-Sb1-LAB3 & 22:17:59.14 & +00:15:28.7 & $103$ & $75\pm9$  & $ 5.2\pm 0.2$ & 0.48 & 1.7 & 3.096$^g$ & X-ray$^l$ \\
SXDS-N-LAB1 & 02:18:21.31 & $-$04:42:33.1 & $101$ & $68\pm5$ & $ 3.3\pm 0.2$ & 0.51 & -0.4 & --- & --- \\
SSA22-Sb1-LAB16 & 22:17:29.01 & +00:07:50.2 & $101$ & $28\pm8$ & $ 0.8\pm 0.2$ & 0.80 & -0.2 & 3.104$^h$ & X-ray$^l$/8$\mu$m$^i$/submm$^n$ \\
 \hline
\end{tabular}
$^a$ Major-axis diameter, $^b$ Filamentarity ($F=0$ for a circle, $F=1$ for a filament, see text for more detail), $^c$\citet{2003ApJ...592..728S}, $^d$this work, $^e$\citet{2007AJ....133.2222S}, $^f$\citet{2002AJ....124.1839B}, $^g$\citet{2005ApJ...634L.125M}, $^h$\citet{2006ApJ...640L.123M}, $^i$\citet{2009ApJ...692.1561W}, $^j$\citet{2001ApJ...548L..17C}, $^k$\citet{2004ApJ...615L..85B}, $^l$\citet{1998AJ....115.1693C}, $^m$\citet{2003AJ....126..539A}, $^n$\citet{2005MNRAS.363.1398G}, $^l$\citet{2009ApJ...700....1G}\\
\end{minipage}
\end{table*}

\begin{figure*}
 \centering
\includegraphics[scale=0.93]{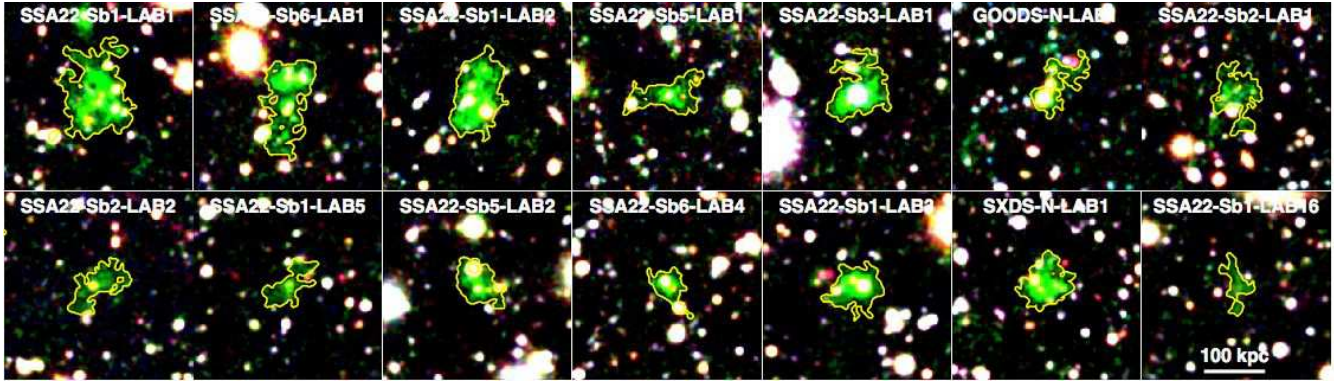} 
  \caption{Pseudo-colour images ($B$ for blue, $NB497$ for green, $V$ for red) of the 14 giant LABs. The size of the images is $40 \times 40$ arcsec$^2$ ($\sim 300 \times300$ kpc$^2$). The yellow contours indicate isophotal apertures with a threshold of $1.4 \times 10^{-18}$ erg s$^{-1}$ cm$^{-2}$ arcsec$^{-2}$. The white horizontal bar in the lower right image represents the angular scale of 100 kpc (physical scale) at $z=3.1$.}
\end{figure*}

We reduced the raw data with {\sc sdfred} \citep{2002AJ....123...66Y, 2003ApJ...582...60O} and {\sc iraf}. We calibrated the astrometry of the images using the 2MASS All-Sky Catalog of Point Sources \citep{2003tmc..book.....C}. For photometric calibration, we used the photometric and spectrophotometric standard stars, SA113, SA115, FEIGE34, Hz44, P177D, GD248, SA95-42, LDS749B, BD+332642, and G24-9 \citep{1990AJ.....99.1621O, 1992AJ....104..340L}. We corrected the magnitudes using the Galactic extinction map of \citet{1998ApJ...500..525S}. We aligned the combined images and smooth with Gaussian kernels to match their seeing to a FWHM of $1.0$ or $1.1$ arcsec depending on the original seeing. We made $BV$ images [$BV=(2B+V)/3$] for the continuum at the same effective wavelength as $NB497$ and made $NB_{\rm c}$ (continuum subtracted $NB497$) images for emission-line images. The total survey area after masking low S/N regions and bright stars is $2.12$ square degrees and the survey volume is $1.6 \times 10^6$ comoving Mpc$^3$. This is 12 times larger than the survey area of \citet{2004AJ....128..569M} and 100 times larger than that of \citet{2000ApJ...532..170S}. The 1-$\sigma$ surface brightness limits of the $NB_{\rm c}$ images are $0.7-1.2 \times 10^{-18}$ erg s$^{-1}$ cm$^{-2}$ arcsec$^{-2}$.

\section[]{Results}

Object detection and photometry are performed using the double image mode of  {\sc SExtractor} version 2.5.0 \citep{1996A&AS..117..393B}. For source detection, we use smoothed $NB_{\rm c}$ images with Gaussian kernels to match their seeing to a FWHM of $1.4$ arcsec in order to slightly increase the sensitivities for diffuse extended sources and to make all the images the same seeing size. We use the same detection threshold (DETECT-THRESH) of $1.4 \times 10^{-18}$ erg s$^{-1}$ cm$^{-2}$ arcsec$^{-2}$ (or 28.5 mag arcsec$^{-2}$) for all the 12 fields. The magnitudes and colours are measured with isophotal apertures defined in the $NB_{\rm c}$ images. 

\begin{figure*}
\centering
\includegraphics[scale=0.52]{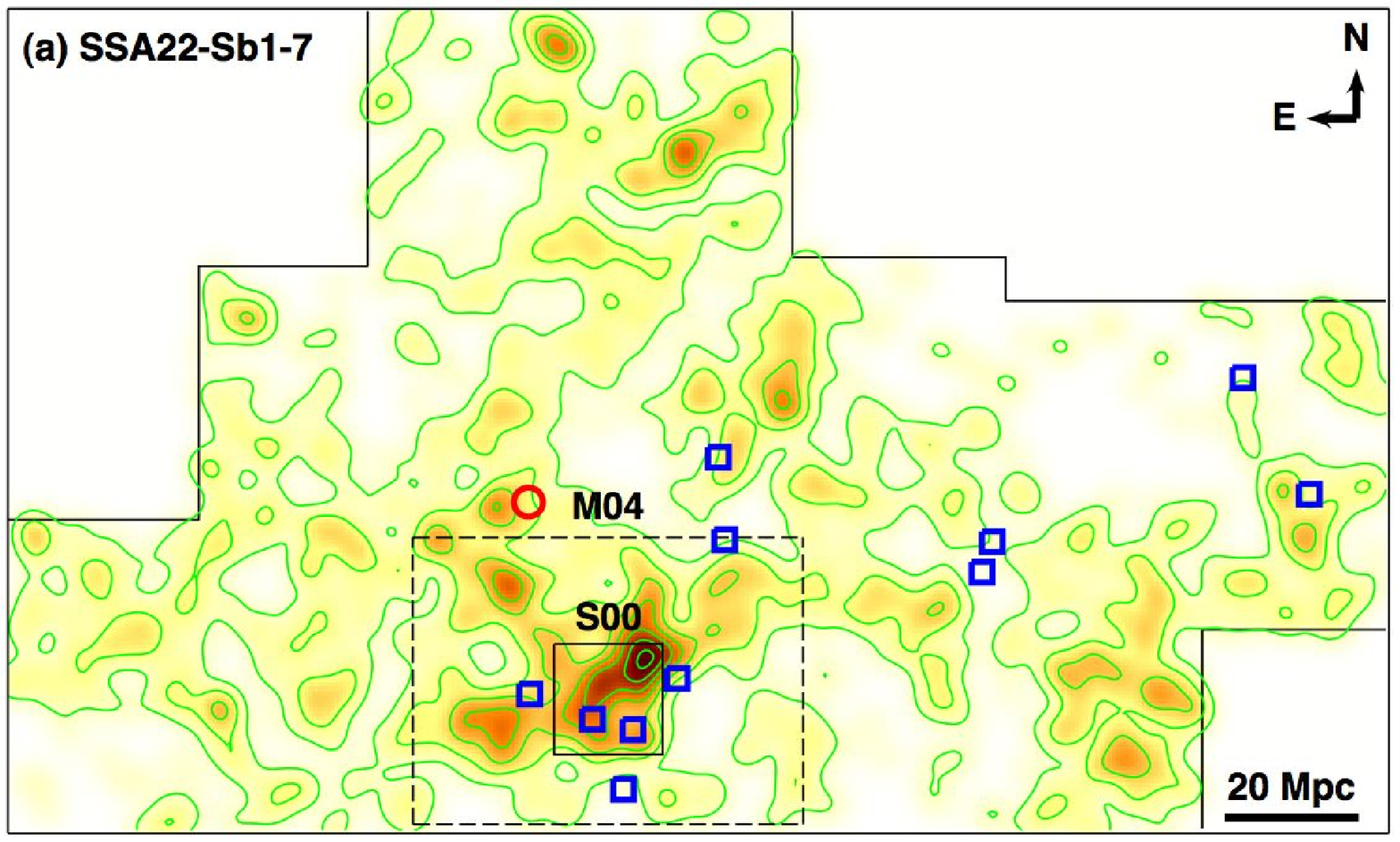}
\hspace{5mm}
\includegraphics[scale=0.26]{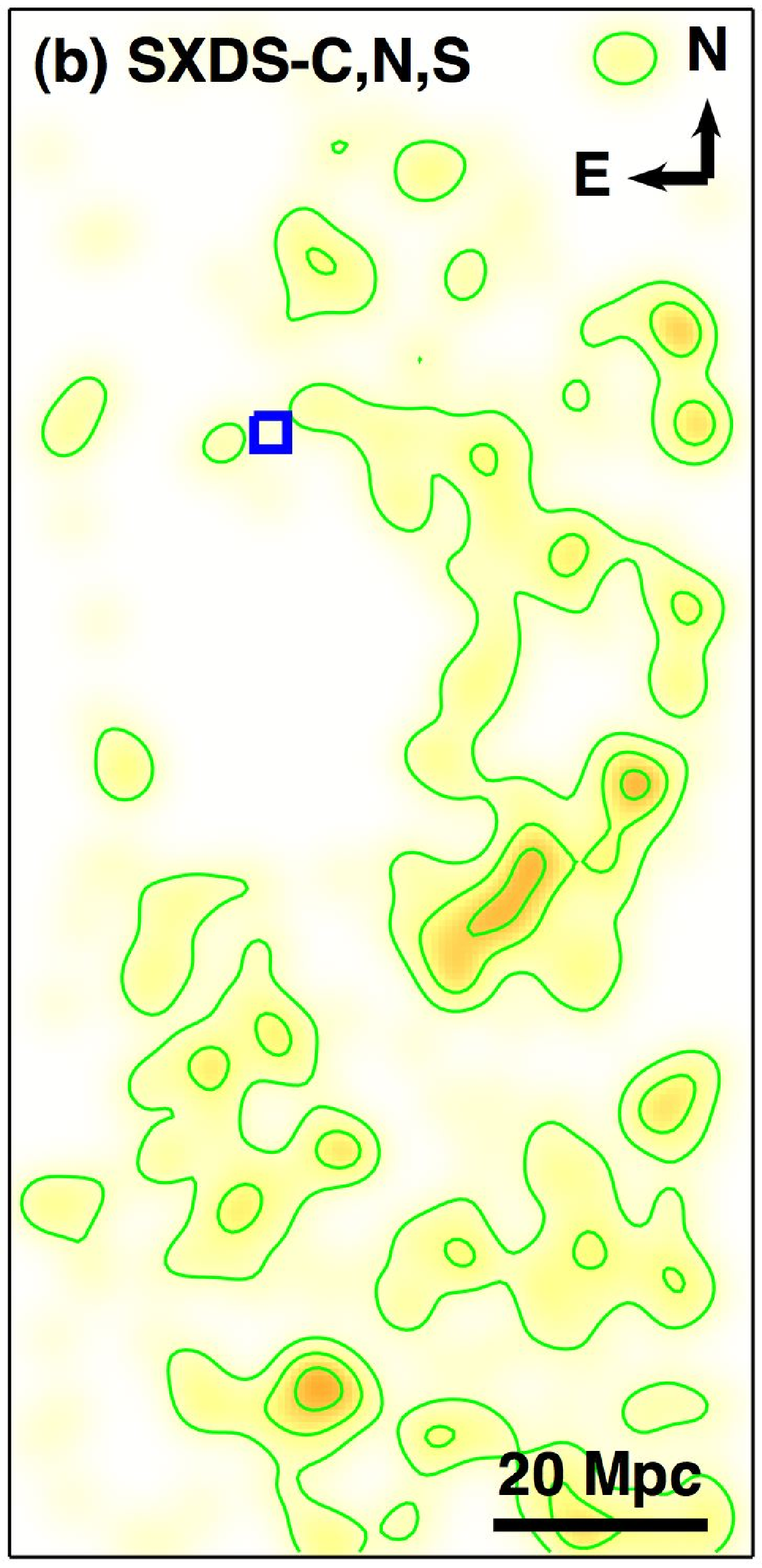} 
\hspace{5mm}
\includegraphics[scale=0.26]{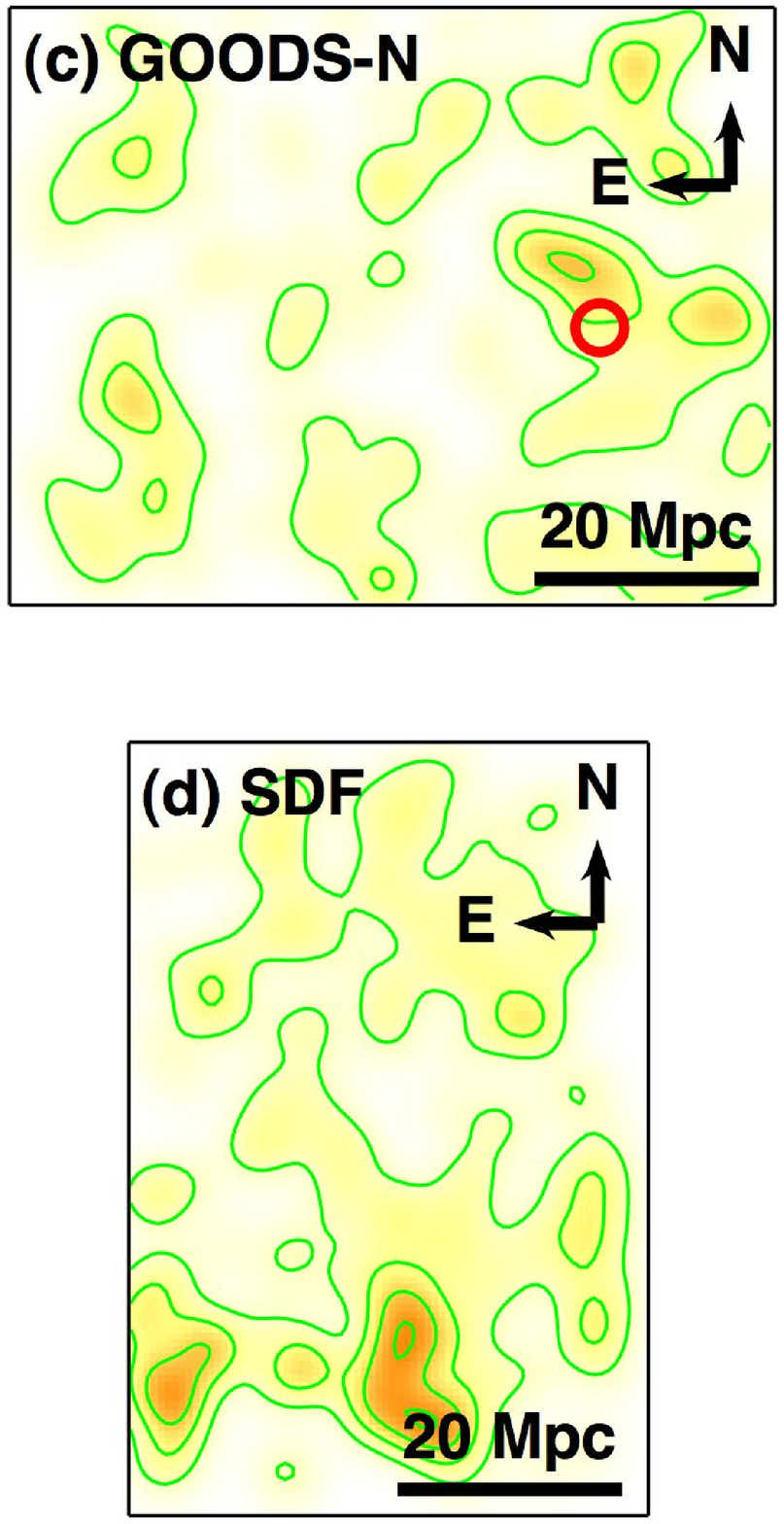} 
  \caption{Sky distribution of the 14 giant LABs and smoothed density maps of $\sim 2000$ compact LAEs at $z\sim 3.09$. In the left panel (a), the small black box indicates SSA22a field by \citet[][S00]{2000ApJ...532..170S} and the dashed box indicates SSA22-Sb1 by \citet[][M04]{2004AJ....128..569M}. The thick bars show the angular scale of 20 comoving Mpc at $z=3.1$. The blue squares and red circles indicate the giant LABs without QSO and with QSO, respectively. The contours represent LAE overdensity, $\delta_{\rm LAE}\equiv(n-\bar{n})/\bar{n}=$0, 1, 2, 3, 4, 5, and 6.}
\end{figure*}

In Fig.~1, we plot the $BV-NB497$ colours and $NB_{\rm c}$ magnitudes of the $NB_{\rm c}$-detected sources. The solid line represents the colour criterion used for narrow-band excess objects, $BV-NB497=0.7$, which corresponds to an observed equivalent width of $EW_{\rm obs}=80$ \AA. From these narrow-band excess objects, we make a diameter-limited catalog of 14 LABs down to a major-axis diameter of the isophotal aperture of $a\ge 13$ arcsec (or $\ge 100$ kpc at $z=3.1$). 

For the LAB selection, we use the major-axis diameters rather than isophotal area, in order to cover LABs with asymmetric structures. For example, cold stream models predicted that LABs have asymmetric, long and thin filaments \citep{2010MNRAS.tmp..933G, 2010arXiv1005.3041F}. An alternative quantity for the LAB selection may be Ly$\alpha$ luminosity. However, the Ly$\alpha$ luminosity could be dominated by a bright central core, such as starbursts in the central galaxy and AGN.

Six out of the 14 LABs have been spectroscopically confirmed by previous surveys \citep{2000ApJ...532..170S, 2002AJ....124.1839B, 2005ApJ...634L.125M, 2006ApJ...640L.123M, 2007AJ....133.2222S}. For two new LABs, we carried out spectroscopic follow-up observations with Magellan/IMACS in June 2009, and identified Ly$\alpha$ emission from these LABs (see Ono et al. in prep for more details).

The properties of the 14 giant LABs are listed in Table~2. We rename the LABs in the new sample since initial surveys. SSA22-Sb1-LAB16 was named as LAB18 in \citet{2004AJ....128..569M}. Note that SSA22-Sb1-LAB1 discovered by \citet{2000ApJ...532..170S} is still the largest one in this sample. SSA22-Sb3-LAB1, and GOODS-N-LAB1 are associated with QSOs \citep{2002AJ....124.1839B, 2007AJ....133.2222S}. We apply the "QSO" labels only to optically bright known QSOs but some of the other LABs have potential signs of obscured AGN. All the 5 LABs from the initial surveys \citep{2000ApJ...532..170S, 2004AJ....128..569M} were detected at X-ray, 8$\mu$m, and/or submm wavelength follow-up observations \citep{2001ApJ...548L..17C, 2004ApJ...615L..85B, 2005MNRAS.363.1398G, 2009ApJ...700....1G, 2009ApJ...692.1561W}. Thus, although there are only two LABs are apparently associated with QSOs, the AGN fraction of the 14 LABs may be higher when accounting for more obscured AGN. 

The thumbnail images ($40\times 40$ arcsec$^2$) of the 14 giant LABs are displayed in Fig.~2. The LABs show a wide variety of Ly$\alpha$ morphologies. While some LABs appear to have circular shapes, some have filamentary (or elongated) shapes. We quantify the Ly$\alpha$ morphology by defining "filamentarity'', \[F \equiv 1 - (({\rm isophotal~ area})/(\pi \times (a/2)^2))\] where $a$ is the major-axis diameter. For example, a circle has $F=0$ and an extremely thin filament has $F=1$. The filamentarities of the LABs range $F \sim 0.4-0.8$. We estimate the uncertainties of the Ly$\alpha$ properties by putting the thumbnail $NB_{\rm c}$ images of each LAB at 100 random positions on the original $NB_{\rm c}$ images and measuring the deviations.

Fig.~3 shows the spatial distributions of the 14 giant LABs and smoothed density maps of $\sim 2000$ Ly$\alpha$ emitters (LAEs) selected by Yamada et al. (in prep) with the same data. Since some bright Ly$\alpha$ knots in the giant LABs are also detected as single or multiple LAEs, we exclude such LAEs from the LAE sample in this analysis. We make the density maps by smoothing the LAE spatial distributions with a Gaussian kernel of $\sigma = 1.2$ arcmin, or FWHM $=5.3$ comoving Mpc. The smoothing kernel size is chosen to match the median distance between the nearest neighbours in the LAE samples in the blank fields (SXDS, GOODS-N, and SDF). The contours represent LAE overdensities, $\delta_{\rm LAE}\equiv(n-\bar{n})/\bar{n}=$0, 1, 2, 3, 4, 5, and 6, where  $\bar{n}$ is the mean LAE number density in the blank fields.

\begin{figure}
\centering
\includegraphics[scale=0.65]{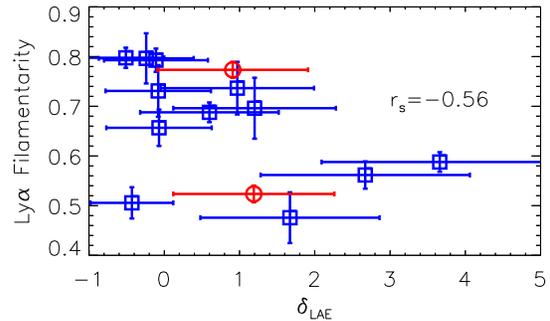}
  \caption{Filamentarity of the 14 giant LABs as a function of the overdensity of LAEs. The blue squares and red circles indicate giant LABs without QSO and with QSO, respectively. The error bars show 1-$\sigma$ uncertainties. The filamentarity of the LABs shows a weak anti-correlation with the overdensity of LAEs.}
\end{figure}

In Fig.~4, we plot the filamentarities of the 14 LABs as a function of the LAE overdensities. The filamentarity shows a weak anti-correlation with the LAE overdensity. While more filamentary LABs reside in average density environments, more circular LABs reside in both average density and overdense environments. The Spearman's rank correlation coefficient is $r_{\rm s}=-0.56$. We can rule out random distributions with $96\%$ confidence. We have confirmed that the results do not change significantly if we use isophotal area for the LAB selection, suggesting that the correlation is not due to the selection method.

\section[]{Discussion and Conclusions}

Based on deep, wide-field Ly$\alpha$ imaging, we construct a sample of 14 giant LAB candidates at $z=3.1$ from a volume of $1.6 \times 10^6$ comoving Mpc$^3$. This is the largest sample of giant LABs and triples the number of known LABs over 100 kpc. Our giant LAB sample shows a wide variety of Ly$\alpha$ morphologies and resides not only in overdense environments, as derived from LAEs, but also in lower dense environments. We find a possible hint for "morphology-density" relation of the LABs: the Ly$\alpha$ filamentarity seems to differ as a function of the local density environments.

How can we interpret this possible morphology-density relation of the LABs? The Ly$\alpha$ morphology may relate to the formation mechanisms of LABs. According to recent numerical simulations, more filamentary LABs may be good candidates for cold gas accretion from the surrounding IGM \citep{2010MNRAS.tmp..933G, 2010arXiv1005.3041F}. Although direct evidence for such gas inflows is not found around star-forming galaxies at $z\sim 2$ \citep{2010arXiv1003.0679S}, recent studies of the metallicity of star-forming galaxies from low- to high-redshifts indicate that gas inflows may still be dominant in the field environment at $z \ga 3$ \citep{2010arXiv1005.0006M}. More circular LABs may relate to large-scale gas outflows, which are driven by intense starbursts and/or AGN activities \citep{2006Natur.440..644M}. At high-redshift, star-formation and AGN activities in overdense environments are known to be several times higher than those in the field environments \citep[e.g.,][]{2003ApJ...599...86S}. Future spectroscopic and multi-wavelength follow-up observations would enable us to investigate the gas dynamics and the variations of the star-formation and AGN activities in giant LABs as a function of the environments and to test the interpretations.

\section*{Acknowledgments}

We thank an anonymous referee for helpful comments which significantly improved the clarity of this paper. We thank Ian Smail for help and useful discussions. YM acknowledges support from STFC. This research is supported in part by the Grant-in-Aid 20540224 for Scientific Research of the Ministry of Education, Science, Culture, and Sports in Japan.

%\appendix

\label{lastpage}


\begin{thebibliography}{99}

\bibitem[\protect\citeauthoryear{Alexander et al.}{2003}]{2003AJ....126..539A} Alexander D.~M., et al., 2003, AJ, 126, 539 

\bibitem[\protect\citeauthoryear{Barger et al.}{2002}]{2002AJ....124.1839B} Barger A.~J., Cowie L.~L., Brandt W.~N., Capak P., Garmire G.~P., Hornschemeier A.~E., Steffen A.~T., Wehner E.~H., 2002, AJ, 124, 1839 

\bibitem[\protect\citeauthoryear{Basu-Zych \& Scharf}{2004}]{2004ApJ...615L..85B} Basu-Zych A., Scharf C., 2004, ApJ, 615, L85 

\bibitem[\protect\citeauthoryear{Bertin \& Arnouts}{1996}]{1996A&AS..117..393B} Bertin, E., \& Arnouts, S.\ 1996, A\&A, 117, 393 

\bibitem[\protect\citeauthoryear{Capak et al.}{2004}]{2004AJ....127..180C} Capak P., et al., 2004, AJ, 127, 180 

\bibitem[\protect\citeauthoryear{Chapman et al.}{2001}]{2001ApJ...548L..17C} Chapman S.~C., Lewis G.~F., Scott D., Richards E., Borys C., Steidel C.~C., Adelberger K.~L., Shapley A.~E., 2001, ApJ, 548, L17 

\bibitem[\protect\citeauthoryear{Condon et al.}{1998}]{1998AJ....115.1693C} Condon J.~J., Cotton W.~D., Greisen E.~W., Yin Q.~F., Perley R.~A., Taylor G.~B., Broderick J.~J., 1998, AJ, 115, 1693 

\bibitem[\protect\citeauthoryear{Cutri et al.}{2003}]{2003tmc..book.....C} Cutri R.~M., et al., 2003, The IRSA 2MASS All-Sky Point Source Catalog, NASA/IPAC Infrared Science Archive, http://irsa.ipac.caltech.edu/applications/Gator/

\bibitem[\protect\citeauthoryear{Dey et al.}{2005}]{2005ApJ...629..654D} Dey A., et al., 2005, ApJ, 629, 654 

\bibitem[\protect\citeauthoryear{Dijkstra \& Loeb}{2009}]{2009MNRAS.400.1109D} Dijkstra M., Loeb A., 2009, MNRAS, 400, 1109 

\bibitem[\protect\citeauthoryear{Faucher-Giguere et al.}{2010}]{2010arXiv1005.3041F} Faucher-Giguere C.~-., Keres D., Dijkstra M., Hernquist L., Zaldarriaga M., 2010, arXiv, arXiv:1005.3041 

\bibitem[\protect\citeauthoryear{Francis et al.}{1996}]{1996ApJ...457..490F} Francis P.~J., et al., 1996, ApJ, 457, 490 

\bibitem[\protect\citeauthoryear{Furlanetto et al.}{2005}]{2005ApJ...622....7F} Furlanetto S.~R., Schaye J., Springel V., Hernquist L., 2005, ApJ, 622, 7 

\bibitem[\protect\citeauthoryear{Furusawa et al.}{2008}]{2008ApJS..176....1F} Furusawa H., et al., 2008, ApJS, 176, 1 

\bibitem[\protect\citeauthoryear{Geach et al.}{2005}]{2005MNRAS.363.1398G} Geach J.~E., et al., 2005, MNRAS, 363, 1398 

\bibitem[\protect\citeauthoryear{Geach et al.}{2009}]{2009ApJ...700....1G} Geach J.~E., et al., 2009, ApJ, 700, 1 

\bibitem[\protect\citeauthoryear{Goerdt et al.}{2010}]{2010MNRAS.tmp..933G} Goerdt T., Dekel A., Sternberg A., Ceverino D., Teyssier R., Primack J.~R., 2010, MNRAS, 933 


\bibitem[\protect\citeauthoryear{Greve et al.}{2007}]{2007MNRAS.382...48G} Greve T.~R., Stern D., Ivison R.~J., De Breuck C., Kov{\'a}cs A., Bertoldi F., 2007, MNRAS, 382, 48 

\bibitem[\protect\citeauthoryear{Haiman, Spaans, \& Quataert}{2000}]{2000ApJ...537L...5H} Haiman Z., Spaans M., Quataert E., 2000, ApJ, 537, L5 

\bibitem[\protect\citeauthoryear{Hayashino et al.}{2004}]{2004AJ....128.2073H} Hayashino T., et al., 2004, AJ, 128, 2073 

\bibitem[\protect\citeauthoryear{Iye et al.}{2004}]{2004PASJ...56..381I} Iye M., et al., 2004, PASJ, 56, 381 

\bibitem[\protect\citeauthoryear{Kashikawa et al.}{2004}]{2004PASJ...56.1011K} Kashikawa N., et al., 2004, PASJ, 56, 1011 

\bibitem[\protect\citeauthoryear{Keel et al.}{1999}]{1999AJ....118.2547K} Keel W.~C., Cohen S.~H., Windhorst R.~A., Waddington I., 1999, AJ, 118, 2547 

\bibitem[\protect\citeauthoryear{Landolt}{1992}]{1992AJ....104..340L} Landolt A.~U., 1992, AJ, 104, 340 

\bibitem[\protect\citeauthoryear{Mannucci et al.}{2010}]{2010arXiv1005.0006M} Mannucci F., Cresci G., Maiolino R., Marconi A., Gnerucci A., 2010, arXiv, arXiv:1005.0006 

\bibitem[\protect\citeauthoryear{Matsuda et al.}{2004}]{2004AJ....128..569M} Matsuda Y., et al., 2004, AJ, 128, 569 

\bibitem[\protect\citeauthoryear{Matsuda et al.}{2005}]{2005ApJ...634L.125M} Matsuda Y., et al., 2005, ApJ, 634, L125 

\bibitem[\protect\citeauthoryear{Matsuda et al.}{2006}]{2006ApJ...640L.123M} Matsuda Y., Yamada T., Hayashino T., Yamauchi R., Nakamura Y., 2006, ApJ, 640, L123 

\bibitem[\protect\citeauthoryear{Matsuda et al.}{2009}]{2009MNRAS.400L..66M} Matsuda Y., et al., 2009, MNRAS, 400, L66 

\bibitem[\protect\citeauthoryear{Miyazaki et al.}{2002}]{2002PASJ...54..833M} Miyazaki S., et al., 2002, PASJ, 54, 833

\bibitem[\protect\citeauthoryear{Mori \& Umemura}{2006}]{2006Natur.440..644M} Mori M., Umemura M., 2006, Natur, 440, 644 

\bibitem[\protect\citeauthoryear{Nilsson et al.}{2006}]{2006A&A...452L..23N} Nilsson K.~K., Fynbo J.~P.~U., M{\o}ller P., Sommer-Larsen J., Ledoux C., 2006, A\&A, 452, L23 

\bibitem[\protect\citeauthoryear{Oke}{1990}]{1990AJ.....99.1621O} Oke, J.~B.\ 1990, AJ, 99, 1621 

\bibitem[\protect\citeauthoryear{Ouchi et al.}{2003}]{2003ApJ...582...60O} Ouchi M., et al., 2003, ApJ, 582, 60 

\bibitem[\protect\citeauthoryear{Palunas et al.}{2004}]{2004ApJ...602..545P} Palunas P., Teplitz H.~I., Francis P.~J., Williger G.~M., Woodgate B.~E., 2004, ApJ, 602, 545 

\bibitem[\protect\citeauthoryear{Prescott et al.}{2008}]{2008ApJ...678L..77P} Prescott M.~K.~M., Kashikawa N., Dey A., Matsuda Y., 2008, ApJ, 678, L77 

\bibitem[\protect\citeauthoryear{Prescott, Dey, \& Jannuzi}{2009}]{2009ApJ...702..554P} Prescott M.~K.~M., Dey A., Jannuzi B.~T., 2009, ApJ, 702, 554 

\bibitem[\protect\citeauthoryear{Schlegel, Finkbeiner, \& Davis}{1998}]{1998ApJ...500..525S} Schlegel D.~J., Finkbeiner D.~P., Davis M., 1998, ApJ, 500, 525 

\bibitem[\protect\citeauthoryear{Shen et al.}{2007}]{2007AJ....133.2222S} Shen Y., et al., 2007, AJ, 133, 2222 

\bibitem[\protect\citeauthoryear{Shimizu \& Umemura}{2010}]{2010MNRAS.406..913S} Shimizu I., Umemura M., 2010, MNRAS, 406, 913 

\bibitem[\protect\citeauthoryear{Smail et al.}{2003}]{2003ApJ...599...86S} Smail I., Scharf C.~A., Ivison R.~J., Stevens J.~A., Bower R.~G., Dunlop J.~S., 2003, ApJ, 599, 86 

\bibitem[\protect\citeauthoryear{Smith \& Jarvis}{2007}]{2007MNRAS.378L..49S} Smith D.~J.~B., Jarvis M.~J., 2007, MNRAS, 378, L49 

\bibitem[\protect\citeauthoryear{Steidel et al.}{2000}]{2000ApJ...532..170S} Steidel, C.~C., Adelberger, K.~L., Shapley, A.~E., Pettini, M., Dickinson, M., \& Giavalisco, M.\ 2000, ApJ, 532, 170 

\bibitem[\protect\citeauthoryear{Steidel et al.}{2003}]{2003ApJ...592..728S} Steidel C.~C., Adelberger K.~L., Shapley A.~E., Pettini M., Dickinson M., Giavalisco M., 2003, ApJ, 592, 728 

\bibitem[\protect\citeauthoryear{Steidel et al.}{2010}]{2010arXiv1003.0679S} Steidel C.~C., Erb D.~K., Shapley A.~E., Pettini M., Reddy N.~A., Bogosavljevi{\'c} M., Rudie G.~C., Rakic O., 2010, arXiv, arXiv:1003.0679 

\bibitem[\protect\citeauthoryear{Taniguchi \& Shioya}{2000}]{2000ApJ...532L..13T} Taniguchi Y., Shioya Y., 2000, ApJ, 532, L13 

\bibitem[\protect\citeauthoryear{Yagi et al.}{2002}]{2002AJ....123...66Y} Yagi M., Kashikawa N., Sekiguchi M., Doi M., Yasuda N., Shimasaku K., Okamura S., 2002, AJ, 123, 66 

\bibitem[\protect\citeauthoryear{Yang et al.}{2009}]{2009ApJ...693.1579Y} Yang Y., Zabludoff A., Tremonti C., Eisenstein D., Dav{\'e} R., 2009, ApJ, 693, 1579 

\bibitem[\protect\citeauthoryear{Webb et al.}{2009}]{2009ApJ...692.1561W} Webb T.~M.~A., Yamada T., Huang J.-S., Ashby M.~L.~N., Matsuda Y., Egami E., Gonzalez M., Hayashimo T., 2009, ApJ, 692, 1561 

\end{thebibliography}
\end{document}